\documentclass[aps,prl,twocolumn,superscriptaddress,showpacs,floatfix]{revtex4}
\usepackage{graphicx}
\usepackage{amsmath}

\newcommand{\dd}{\ensuremath{\text{d}}} 

\newcommand{\Eu}{\ensuremath{\text{e}}}
\newcommand{\kb}{\ensuremath{k_\text{B}}}
\newcommand{\dds}[1]{\frac{\partial}{\partial #1}}

\newcommand{\Eq}[1]{Eq.~(\ref{eqn:#1})}
\newcommand{\Fig}[1]{Fig.~\ref{fig:#1}} 
\newcommand{\Tab}[1]{Table \ref{table:#1}}
\newcommand{\Na}[1]{\ensuremath{^{#1}}\text{Na}}

\begin{document}
\title{Enhanced heat flow in the  hydrodynamic-collisionless regime}
\author{R.~Meppelink, R.~van Rooij, J.~M.~Vogels and P.~van der Straten}
\affiliation{Atom Optics and Ultrafast Dynamics, Utrecht University,\\ P.O. Box 80,000, 3508 TA Utrecht, The Netherlands}
\date{\today}

\begin{abstract} 

We study the heat conduction of a cold, thermal cloud in a highly asymmetric trap. The cloud is axially hydrodynamic, but due to the asymmetric trap radially collisionless. By locally heating the cloud we excite a thermal dipole mode and measure its oscillation frequency and damping rate. We find an unexpectedly large heat conduction compared to the homogeneous case. The enhanced heat conduction in this regime is partially caused by atoms with a high angular momentum spiraling in trajectories around the core of the cloud. Since atoms in these trajectories are almost collisionless they strongly contribute to the heat transfer. We observe a second, oscillating hydrodynamic mode, which we identify as a standing wave sound mode.
\end{abstract}

\maketitle
The field of Bose-Einstein condensation in dilute atomic gases provides a fruitful playground to test well-developed theories of quantum fluids. Research using Bose-Einstein condensates (BECs) can address open questions relating to the many-body aspects of two-component quantum liquids, namely the interaction between the hydrodynamic normal and the superfluid component at finite temperatures \cite{JLowTempPhys.116.277}. After the first realization of BEC some pilot experiments have been carried out, but detailed experiments are missing \cite{PhysRevLett.78.764,PhysRevLett.81.500}. This has to be compared to the case of liquid helium below the $\lambda$ point, where many experiments since the 1950s have added to our understanding of novel phenomena in quantum liquids, like collective excitations, first and second sound, and others.  One of the drawbacks of liquid helium is that the interactions are so strong that a clear distinction between the two components is difficult.

The reason for the lack of detailed experiments in BECs to study quantum liquids and in particular the hydrodynamical aspects of it, is the limited number of atoms (typically 1--10 million) in the experiments leaving the thermal atoms virtually collisionless. Efforts to decrease the mean free path by increasing the confinement limits the lifetime of the sample, since the density is limited by three-body decay. This makes the observation of sound propagation in a BEC a challenge. 

As to theory, hydrodynamical damping of trapped Bose gases has been described above and below the transition temperature $T_\text{c}$ \cite{JLowTempPhys.111.793,JLowTempPhys.116.277}. These theories yield the oscillation frequencies and damping rates of several low-lying modes, where it is assumed that the sample is fully hydrodynamic in all directions. Experiments on dilute clouds of cold atoms are generally conducted in highly asymmetric traps. In these elongated, cigar-shaped geometries the mean free path of the atoms can become much shorter than the size of the cloud in the long, axial direction, but at the same time exceeds the size in the other, radial directions. In this so-called hydrodynamic-collisionless regime the system is axially hydrodynamic and radially collisionless. In our setup, described in detail in Ref.~\cite{RevSciInstr.78.013102}, we have created BECs containing up to 3 $\times$ 10$^8$ sodium atoms by evaporation of atoms in an axially strongly decompressed trap with an aspect ratio of 1:65. Hot atoms created in three-body collisions are able to leave the sample in this highly asymmetric trap, before they can heat other atoms in an avalanche \cite{PhysRevA.75.031602}. The sample is axially hydrodynamic, but due to the large aspect ratio collisionless in the radial direction. Such samples seem ideal for the observation of sound propagation in the axial direction, since the axial length of the condensates exceeds a few mm. However, neither experiments nor theoretical descriptions exist to determine if the collisions in the radial direction will affect the damping rates to a degree that the observation of sound remains elusive.

From a practical point of view, the hydrodynamic-collisionless regime is of relevance for the realization of a continuous atom laser by evaporatively cooling a magnetically guided atom beam \cite{PhysRevLett.93.093003}. Here the efficiency of the cooling process is expected to be limited by the heat transfer between the hot, upstream and cold, downstream parts of the beam. 

In this Letter we report the experimental determination of the heat conduction in a cold, thermal gas above the transition temperature $T_\text{c}$, which is hydrodynamic in the axial direction, but collisionless in the radial directions. The heat conduction is determined by locally heating the cloud and subsequently observing the equilibration of the temperature distribution. Two previously unobserved hydrodynamic modes are reported; a thermal dipole mode and a standing wave sound mode. Furthermore, by reducing the number of atoms the transition in the axial direction from the hydrodynamic regime to the collisionless regime is observed. We find that the heat conduction is five times stronger than calculations for the homogeneous case predict.

\begin{figure*}[!ht]
  \begin{center}
    \includegraphics[width=0.65\textwidth]{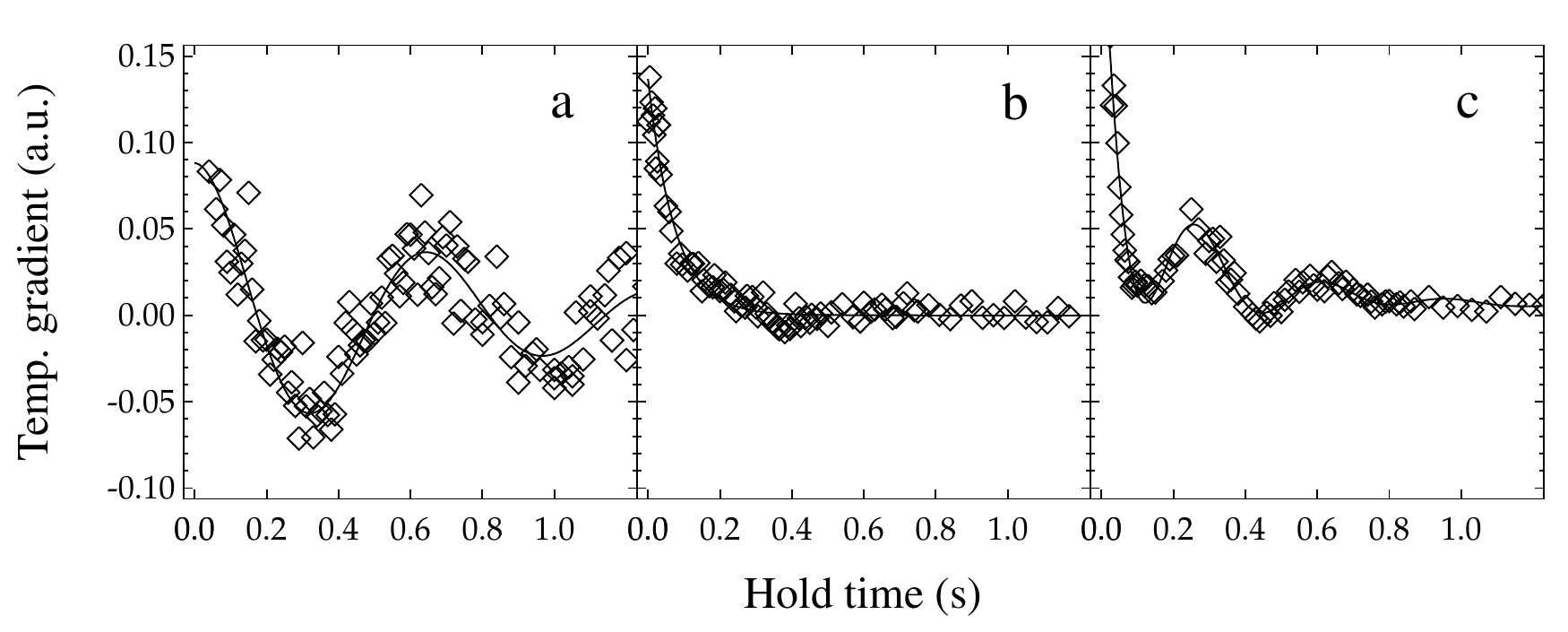}
  \end{center}
  \caption{The temperature gradient as a function of the hold time $\tau$ for three values of the hydrodynamicity parameter $\bar{\gamma}$: $\bar{\gamma}\approx 1$ (a), $\bar{\gamma}\approx 3$ (b), $\bar{\gamma}\approx 7.5$ (c). In figure (a) the behavior is collisionless, where the heat propagates through the MT almost undamped. In Fig. (b) the behavior is neither collisionless nor hydrodynamic and the temperature gradient is nearly critically damped. In Fig. (c) the behavior is hydrodynamic and the damping rate is therefore lower compared to (b), but a distinctive oscillatory behavior can be identified. Due to the destructive imaging scheme used each point represents the temperature gradient of a newly prepared cloud.}
  \label{fig:thermcond-coll}
\end{figure*}

We measure the heat flow by locally heating the thermal cloud, after which the equilibration is studied. The heat is induced by exciting the thermal cloud using Bragg scattering with a laser beam aligned perpendicular to the axial axis of the cloud, which is aimed at its tail and retro-reflected  \cite{PhysRevLett.82.871}. The ($1/\Eu$)-width of the intensity of this beam is $0.8$ mm, which is close to half of the axial ($1/\Eu$)-size of the cloud. The asymmetric excitation is chosen since it yields the maximum separation between the cold, unperturbed part and the heated part of the cloud, resulting in a long observation time. The laser beam is detuned 2 nm below of the $\Na{23} \, \text{D}_2$ transition in order to reduce resonant scattering and prevent superradiant scattering. The excited particles will locally redistribute their momentum and energy through collisions with the other particles, resulting after a few collisions in a \it local \rm thermal equilibrium. 

The experiments are conducted on a cloud containing up to $1.3 \times 10^9$ atoms confined in a clover leaf type magnetic trap (MT) characterized by the radial trap frequencies $\omega_\text{rad}=2 \pi\times 95.6$ Hz and the axial trap frequency $\omega_\text{ax}=2 \pi \times 1.46$ Hz at a temperature between 1.2 and 2 $\mu$K, which is above $T_\text{c} \approx 1$ $\mu$K.  Once a cloud is excited, it is allowed to rethermalize in the MT during an adjustable hold time $\tau$, after which the confinement is turned off and an absorption image is taken after time-of-flight (TOF). The TOF duration is chosen in such a way that the optical density will not exceed 3.5, resulting in a time-of-flight of 40 ms for the highest number of atoms.


We introduce a measure for the hydrodynamicity in the axial direction $\bar{\gamma}\equiv \gamma_\text{col}/\omega_\text{ax}$, where the collision rate $\gamma_\text{col} = n_\text{eff} \,\sigma\, v_\text{rel}$ is the average number of collisions. Here, the relative velocity $v_\text{rel}=\sqrt{2} \bar{v}_\text{th}$, where $\bar{v}_\text{th}=\sqrt{8\kb T/m \pi}$ is the thermal velocity at temperature $T$ and $m$ is the mass and $\sigma =8\pi a^{2}$ is the isotropic cross section of two bosons with s-wave scattering length $a$. Furthermore, $n_\text{eff}=\int n^2\left (\vec{r}\right )\dd V/\int n\left(\vec{r}\right )\dd V = n_0/\sqrt{8}$ for an equilibrium distribution in a harmonic potential, where $n_0$ is the peak density. Written in terms of the number of atoms  $N=n_0 \left ( 2 \pi \kb T/ \left ( m \bar{\omega}^2  \right ) \right )^{3/2}$ and the geometric mean of the angular trap frequencies $\bar{\omega}^3 \equiv \omega^2_\text{rad} \omega_\text{ax}$, this results in $\gamma_\text{col}= N m \sigma \bar{\omega}^3/(2 \pi^2 \kb T) \approx 90 \text{ s}^{-1}$ for the highest number of atoms and corresponds to a hydrodynamicity of  $\bar{\gamma}\lesssim$ 10 in the axial direction. Even at the highest hydrodynamicity the lifetime of the cloud, limited by 3-body decay, is more than 60 s, which is more than sufficient for the experiments described below. Note that the hydrodynamicity parameter in the radial direction is due to the anisotropic trap potential $\bar{\gamma}_\text{rad}\lesssim 10/65$ and the cloud is in the radially collisionless regime. By reducing the number of atoms the heat flow through the thermal cloud can also be measured in the axially collisionless regime.

The images are analyzed using a least square fit to a 2D Gaussian distribution. In this distribution the radial size as a function of the axial position is modeled by a hyperbolic tangent function which adds a gradient to the width that resembles the asymmetric distribution. The fit to the 2D distribution yields the temperature gradient, axial and radial cloud sizes, and the optical density. In the following we will focus on the temperature gradient, which is a measure of the imbalance of the temperature in the cloud. In  this analysis we assume that a local temperature equilibrium is established at all times. Since this is not a valid assumption in the first few tenths of milliseconds after excitation especially for lower collision rates, we cannot accurately describe the data at these times. 

Heating the thermal cloud will also cause a quadrupole motion of the atoms, since the cloud is excited non-adiabatically to a higher temperature. Since the resulting compression and decompression is homogeneous over the cloud it does not influence the temperature gradient. The quadrupole mode, induced by perturbing the magnetic confinement, has been studied experimentally by Buggle and coworkers \cite{PhysRevA.72.043610}. They found the damping rate and oscillation frequency of this mode to be in good agreement with a theoretical model \cite{PhysRevA.60.4851}. Since the quadrupole mode damps slower than the thermal dipole mode considered in this paper, the maximum hold time for most of our measurement series turns out to be insufficient to accurately determine the damping rate of the quadrupole mode. Although our results are less accurate, we have confirmed that the frequency and damping rate of the quadrupole mode are in agreement with the results reported in Refs.~\cite{PhysRevA.72.043610,PhysRevA.60.4851}.

A series of measurements consists of about 100 shots at various hold time $\tau$ from which the temperature gradient is determined. The number of atoms and temperature is determined from an average over all shots, which yields the hydrodynamicity $\bar{\gamma}$. We plot the temperature gradient as a function of $\tau$ for three values of $\bar{\gamma}$ in \Fig{thermcond-coll}.

\begin{figure}[!ht]
  \begin{center}
    \includegraphics[width=0.4\textwidth]{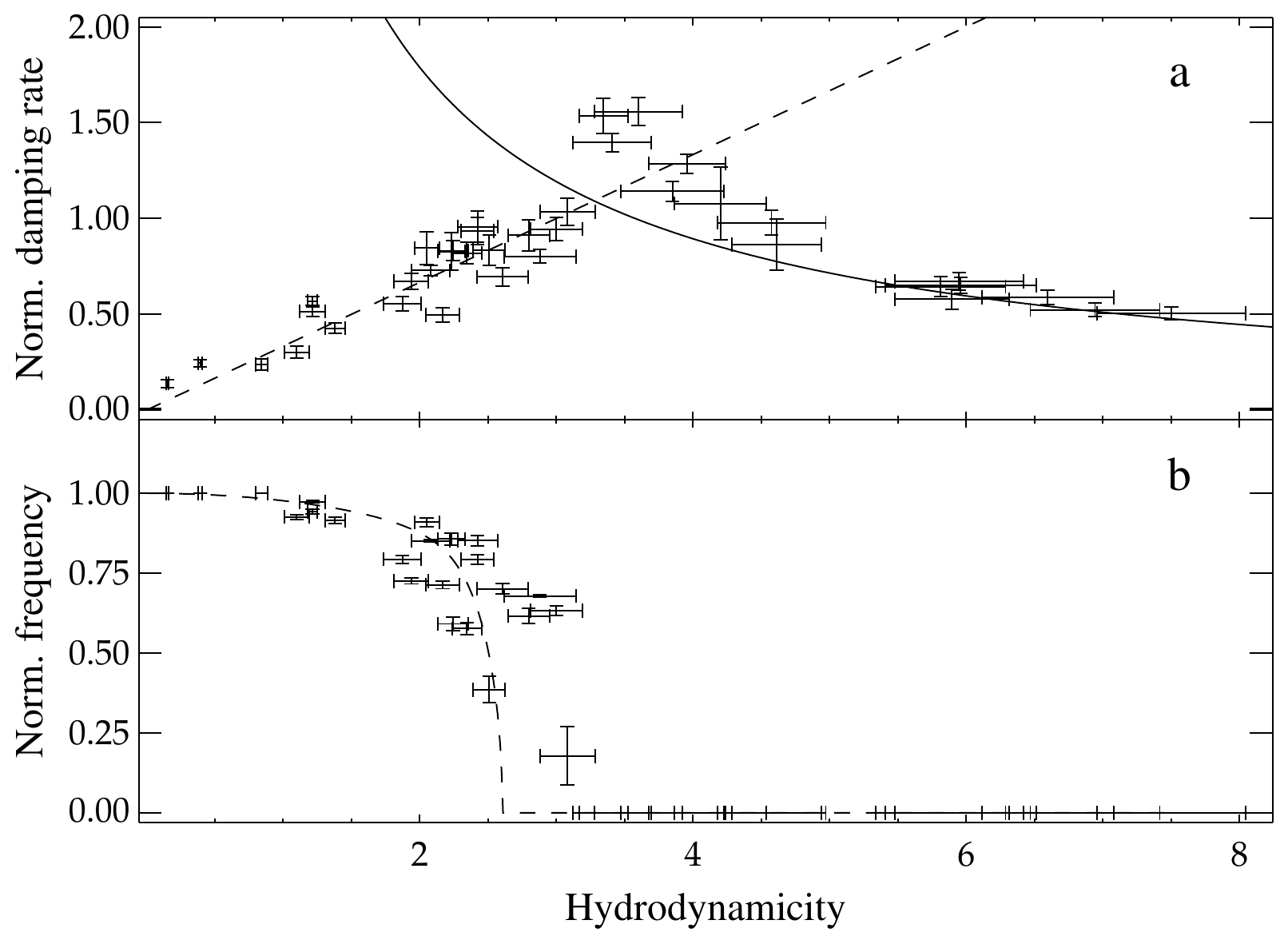}
  \end{center}
  \caption{The measured normalized damping rate $\Gamma_\text{d}/\omega_\text{ax}$ (a) and normalized frequency $\omega_\text{d}/\omega_\text{ax}$ (b) of the thermal dipole mode as a function of the hydrodynamicity $\bar{\gamma}$. The solid line in (a) is a fit of the data points with $\bar{\gamma}>5$ to the solution of \Eq{heatdiffusion} with $\kappa_0=6.4$. The dashed lines are a guide to the eye. The vertical error bars show only statistical errors; the main contribution to the uncertainty in $\bar{\gamma}$ is the uncertainty in the number of atoms.}
  \label{fig:thermal-dipole}
\end{figure}

\Fig{thermcond-coll}(a) is the result of a measurement at small $\bar{\gamma}$ and shows a slowly damped oscillation, where the temperature gradient after half a trap oscillation has changed sign. The heated atoms are then at the opposite side of the cloud with respect to the excitation side, oscillating a frequency $\omega_\text{d}/ \omega_\text{ax}=1$. We refer to this mode as the thermal dipole mode, which has not been observed previously. The oscillation frequency $\omega_\text{d}$ and damping rate $\Gamma_\text{d}$ are determined by fitting the data to a damped sinusoidal and are shown in \Fig{thermal-dipole} as a function of $\bar{\gamma}$. For small collision rates ($\bar{\gamma}\approx1$) the damping rate is proportional to the collision rate. As a consequence the frequency of the mode will decrease for increasing $\bar{\gamma}$ until it reaches zero, when the system is critically damped. In the experiment we observe oscillatory behavior for $\bar{\gamma} \lesssim 2.5$.

For higher values of $\bar{\gamma}$ the temperature gradient as a function of $\tau$ becomes critically damped, as can be seen in \Fig{thermcond-coll}(b). For even higher values of $\bar{\gamma}$ the system becomes hydrodynamic and atoms cannot move through the cloud without colliding frequently. The heat transport will become diffusive, which is a slower process than the harmonic oscillation. As a consequence we expect the damping of the temperature gradient to decrease, but remain non-oscillatory. A measurement for high $\bar{\gamma}$ is shown in \Fig{thermcond-coll}(c). For $\tau<0.1$ s, a double exponential decay can be seen, where the fast decay due to higher order modes and the slow decay of the lowest thermal dipole mode can be discriminated from each other due to the strong inequality of the damping rates. The reduced chi-squared for all fits are of the order of unity, as can also be seen from \Fig{thermcond-coll}, since the curves go smoothly through the data points.

The measurements in the hydrodynamic regime can be analyzed by numerically solving the heat diffusion equation in the axial direction
\begin{equation}
  c_p n(z) \dds{t} T(z,t) = \dds{z}\left[\kappa \dds{z} T(z,t)\right].
  \label{eqn:heatdiffusion}
\end{equation}
Here the specific heat capacity $c_p=7/2$, the heat conductivity $\kappa \equiv \kappa_0 \pi v_\text{th} \Sigma/(\sqrt{8}\sigma)$ with $\kappa_0$ the dimensionless heat conductivity coefficient and $\Sigma=2 \pi \kb T/(m \omega_\text{rad}^2)$ is the effective cloud surface.  The damping rate of the lowest order solution in this regime is found to be $\Gamma_\text{d}=0.542 \kappa_0/\bar{\gamma}$. Fitting the measurements for $\Gamma_\text{d}$ with $\bar{\gamma}>5$ yields $\kappa_0=${6.4}$\pm${0.4}. This value is a factor of five larger than the Chapman-Enskog value $\kappa_0 =75/64 \approx 1.17$ for a homogeneous hydrodynamic system \cite{Chapman1916,Enskog1917}.

\begin{figure}[!ht]
  \begin{center}
    \includegraphics[width=0.4\textwidth]{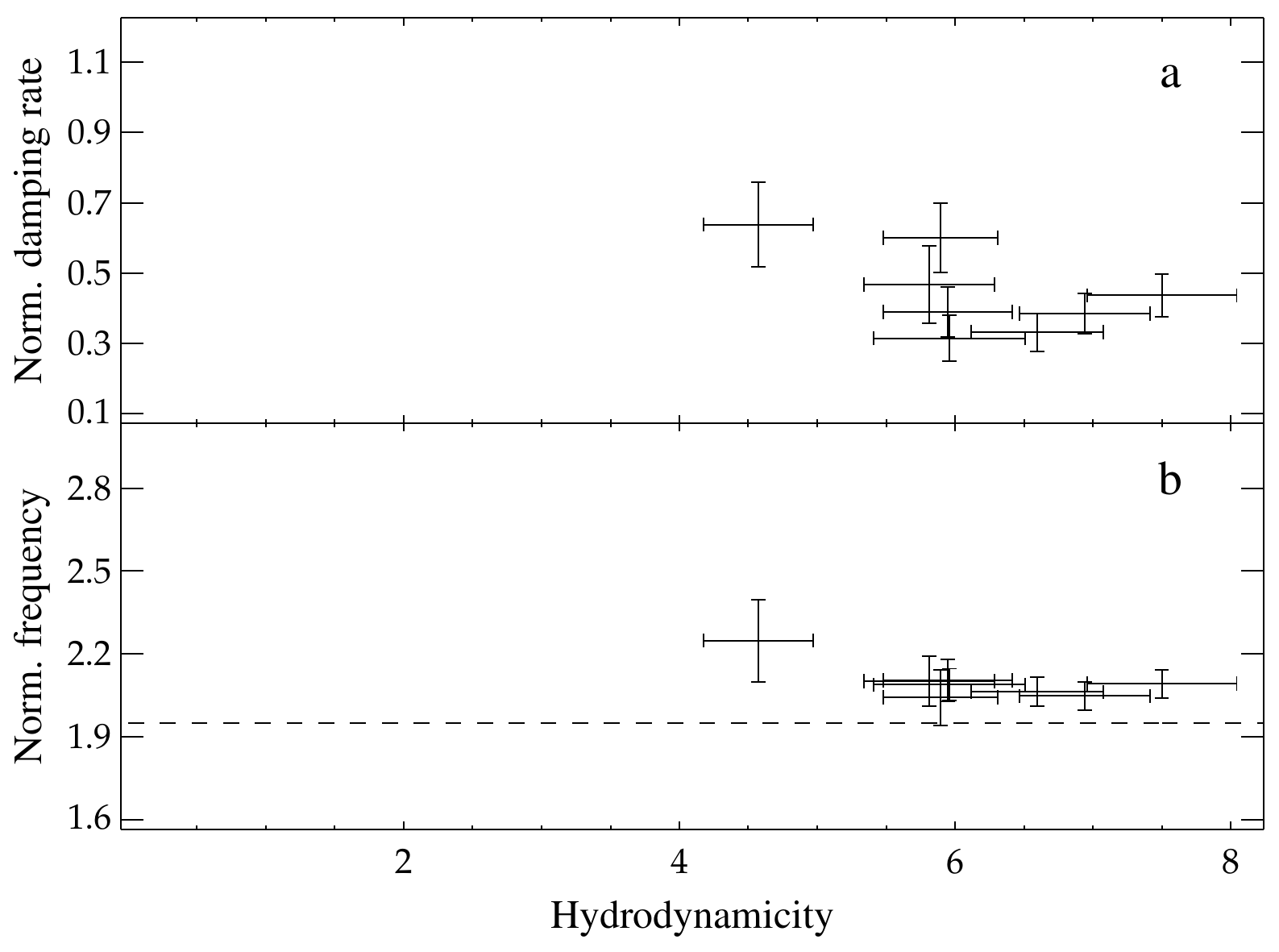}
  \end{center}
  \caption{The measured normalized damping rate $\Gamma_\text{s}/\omega_\text{ax}$ (a) and normalized frequency $\omega_\text{s}/\omega_\text{ax}$ (b) of the hydrodynamic sound mode as a function of the hydrodynamicity $\bar{\gamma}$. The dashed line shows the frequency of this mode in the limit of no damping, $\omega=\sqrt{19/5}$.}
  \label{fig:thermal-sound}
\end{figure}

The measurements in the hydrodynamic regime also show a damped oscillation of the decay of the temperature gradient for $\tau > 0.1$ s (\Fig{thermcond-coll}(c)). As this oscillation is only seen in the hydrodynamic regime, where the collisionless oscillation is completely damped out, we conclude that it is the result of another hydrodynamic mode, which we identify as a standing wave sound mode. In our experiments this sound mode can only be seen for values of $\bar{\gamma}$ exceeding 5. Using a least-square fit we determine both the oscillation frequency $\omega_\text{s}$ and damping rate $\Gamma_\text{s}$ of the sound mode for all values of $\bar{\gamma}>5$ as a function of $\bar{\gamma}$ (see \Fig{thermal-sound}). The measured normalized frequency $\omega_\text{s}/\omega_\text{ax} \approx 2.1$ confirms that the mode differs from a center-of-mass motion $\omega_\text{ax}$ and the quadrupole mode $\omega_\text{q}/\omega_\text{ax} \approx \sqrt{12/5}$ \cite{PhysRevA.60.4851}.

\begin{figure}[htbp]
\begin{center}
  $\begin{tabular}{cc}
  \includegraphics[width=0.22\textwidth]{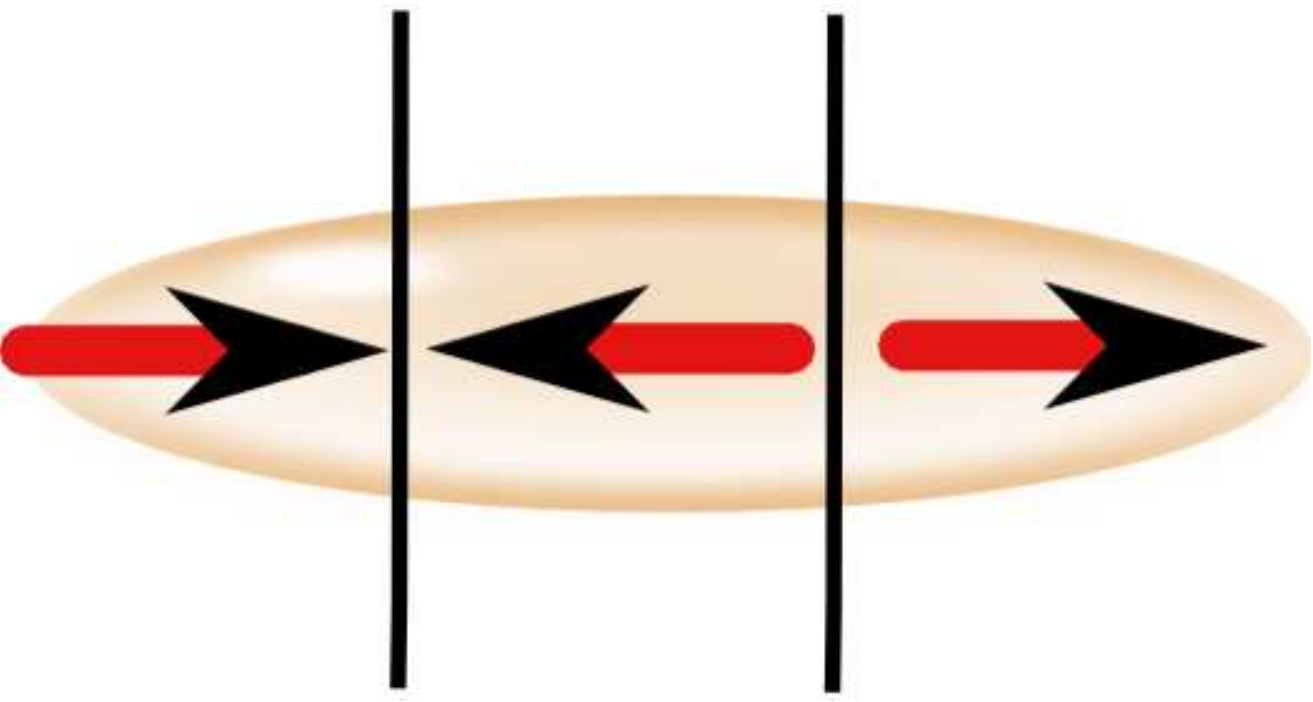} &  \includegraphics[width=0.22\textwidth]{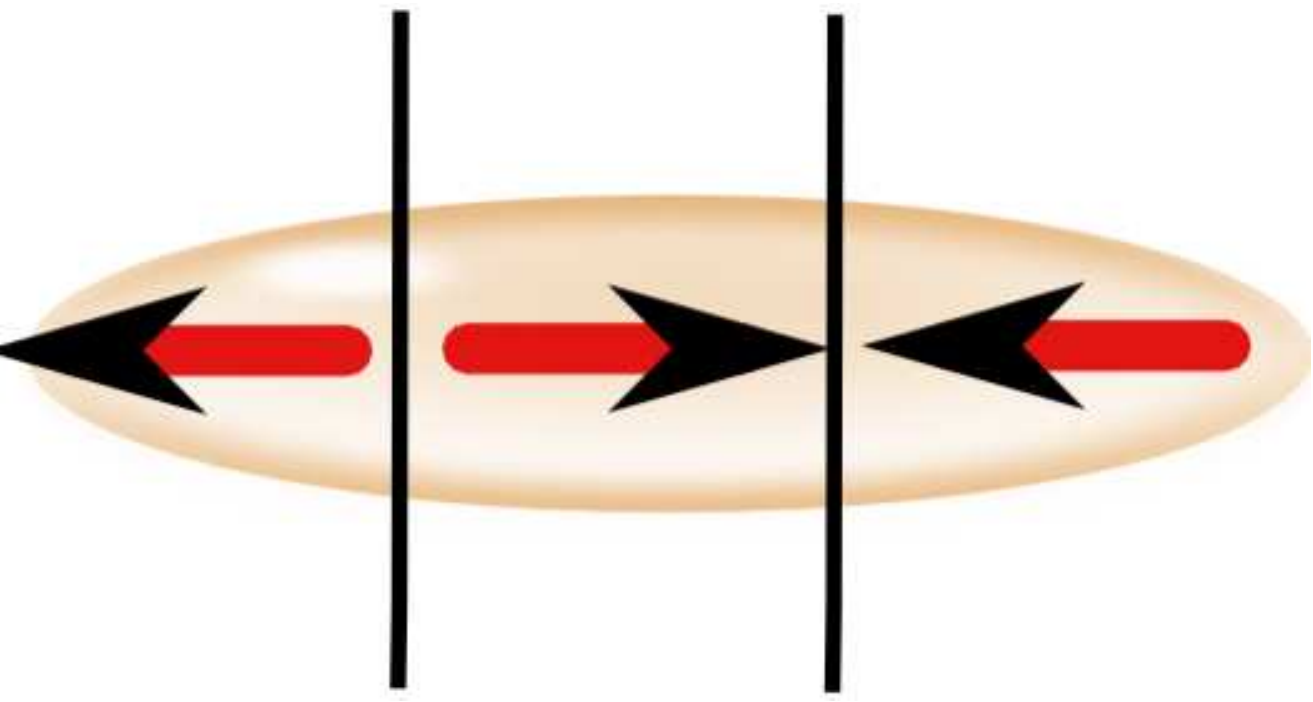} \\
\end{tabular}$
\end{center}
\caption[sound]
{Schematic representation of the oscillating, higher order hydrodynamic mode, a standing wave sound mode. The vertical lines indicate the velocity nodes, the arrows indicate the motion of the atoms.}
\label{fig:sound}
\end{figure}

This sound mode can be found theoretically by solving the hydrodynamic equations \cite{JLowTempPhys.111.793} in the limit of no damping and we find $\omega=\sqrt{19/5}\omega_\text{ax}\approx 1.95\omega_\text{ax}$. This mode resembles the quadrupole mode, although the standing wave sound mode is even in the axial velocity $v_z$ and has two nodes in the velocity profile instead of one. A schematic representation of this mode is given in \Fig{sound}.  As a consequence, this sound mode contributes to the temperature gradient and can be observed in \Fig{thermcond-coll}(c). This is the first direct experimental observation of a thermal sound mode in a cold gas. 

A rigorous theoretical model to calculate the oscillation frequency and damping rate of the modes for arbitrary hydrodynamicity $\bar{\gamma}$ will be presented in Ref. \cite{danismartien10}. The analysis yields $\kappa_0 = 5.98$ in the hydrodynamic-collisionless regime, which confirms the experiment. Furthermore, for a linear confinement as is used for magnetically guided atomic beams the enhancement of the heat flow is even stronger; up to two orders of magnitude. This result implies that the efficiency of evaporatively cooling of a linearly guided atomic beam is strongly diminished due to the large heat flow and questions the feasibility of realizing a continuous atom laser. The increase of $\kappa_0$ is found to be caused by two effects. First, the shape of the density of states is altered, which results in a relatively lower collision rate and causes the presence of more atoms with a high transverse energy. The second effect lies in the presence of atoms with a high angular momentum. These atoms are in trajectories spiraling around the core of the cloud. These trajectories are almost collisionless and cause a very strong contribution to the heat flow.

In conclusion, we have successfully excited and measured two previously unobserved modes; a thermal dipole mode and a standing wave sound mode. Observation of the latter demonstrates both the hydrodynamic behavior of the cloud and the presence of sound propagation in a dilute thermal gas. In the hydrodynamic regime we have measured the heat conduction coefficient $\kappa_0=${6.4}$\pm${0.4}, which is five times higher than calculations for the homogeneous case predict. This effect is expected to be even stronger for a linear confined cloud in the axially hydrodynamic-radially collisionless regime. This result implies that the efficiency of evaporative cooling in a continuous atom laser, where the confinement is linear, is strongly reduced by the large heat transfer between the hot and cold parts of the beam.

This work is supported by the Stichting voor Fundamenteel Onderzoek der Materie ``FOM'' and by the Nederlandse Organisatie voor Wetenschaplijk Onderzoek ``NWO''. We are grateful to W.~C.~Germs for contributing in the early stages to the theoretical description and to D.~Gu\'ery-Odelin and J.~Dalibard for stimulating discussions.


\end{document}